\documentclass[11pt]{article}
\usepackage[toc,page]{appendix}
\RequirePackage{fix-cm}
\usepackage[utf8]{inputenc}
\usepackage[margin=1.5in]{geometry}
\usepackage[bottom, flushmargin]{footmisc}
\usepackage{marvosym}
\usepackage{titlesec}

\titlespacing\section{0pt}{0.21in}{0.03in}
\titlespacing\subsection{0pt}{0.21in}{0.01in}
\titlespacing\subsubsection{0pt}{0.21in}{0in}
\usepackage{marginnote}

\usepackage[superscript]{cite}
\usepackage{enumitem} 
\newcommand{\bulletlabel}{\raisebox{0.2ex}{\small$\bullet$}}
\newlist{enumbul}{enumerate}{6}
\setlist[enumbul]{label=\bulletlabel, topsep=0pt, leftmargin=0.3in, rightmargin=0.3in}
\newlist{enumar}{enumerate}{6}
\setlist[enumar]{label=\arabic*$)$,topsep=0pt, leftmargin=0.3in, rightmargin=0.3in}

\usepackage{hyperref}
\usepackage{comment} % commenting out a large chunk
\usepackage{enumitem} % lists
\usepackage{graphicx,nicefrac} % graphics, especially external ones
\graphicspath{ {dataviz/} }
\usepackage{amsmath} % writing math
\usepackage{amssymb} % writing symbols
\usepackage{amsthm}
\usepackage{bm}
\usepackage{float}
\usepackage{mathtools}
\usepackage{bigints}

\usepackage[most]{tcolorbox}
\usepackage{color} % colors
\usepackage{xcolor, colortbl} % more control over colors

\hypersetup{
	colorlinks,
	linkcolor=black, %internal links
	urlcolor=cyan,  % URLs
	citecolor= red, % references
	breaklinks=true
}

\usepackage{multirow}
\usepackage{multicol}
\usepackage{array}
\usepackage{booktabs}
\newcolumntype{L}[1]{>{\raggedright\let\newline\\\arraybackslash\hspace{0pt}}m{#1}}
\newcolumntype{C}[1]{>{\centering\let\newline\\\arraybackslash\hspace{0pt}}m{#1}}
\newcolumntype{R}[1]{>{\raggedleft\let\newline\\\arraybackslash\hspace{0pt}}m{#1}}

\def\highlight#1#2{
	\begin{centering}
		\begin{minipage}{1\linewidth}
			\begin{tcolorbox}[boxsep = 3.3pt, boxrule=0.4pt, colback = black!4, colbacktitle=black!4, coltitle = black, colframe= black, title=\centering \textsc{#1}]
				\centering
				#2
			\end{tcolorbox}
		\end{minipage}		
	\end{centering}
}

\begin{document}
\pagenumbering{gobble}
\def\acronymend{MECo}
\def\acronym{\acronymend\space}

\def\TOTALBALLOTS{25,545}
\def\TOTALCONTESTS{9,704}
\def\TOTALPARTIES{110}
\def\TOTALCOALITIONS{18}
\def\TOTALCONTESTSSTATE{6,936}
\def\TOTALCONTESTSFEDERAL{2,715}
\def\TOTALCONTESTSBYELEC{53}
\def\TOTALELECTIONS{16}
\def\UNIQUECANDIDATES{14,000}
\def\TOTALCORRECTIONS{114}

\highlight{Accepted for Publication}{
	This version of the article has been peer-reviewed and accepted for publication in \textit{Scientific Data}. However, this is not the Version of Record and does not reflect post-acceptance improvements, or any corrections. The Version of Record is available at: \href{https://doi.org/10.1038/s41597-025-06502-7}{doi.org/10.1038/s41597-025-06502-7}
}

\vfill

\begin{center}
	\begin{minipage}{0.88\linewidth}
		\begin{center}
			\LARGE
			The Malaysian Election Corpus (\acronymend):\\Federal and State-Level Election Results\\from 1955 to 2025\\[0.21in]
			\Large
			Thevesh Thevananthan\footnotemark$^{\text{\Letter}}$\\[0.15in]
		\end{center}
		\large
		Empirical research and public knowledge on Malaysia's elections have long been constrained by a lack of high-quality open data, particularly in the absence of a Freedom of Information framework. This paper introduces the Malaysian Election Corpus (MECo), an open-access panel database covering all federal and state general elections since 1955, as well as by-elections since 2008. MECo includes candidate- and constituency-level data for 9,704 electoral contests across seven decades, standardised with unique identifiers for candidates, parties, and coalitions. The database also provides summary statistics for each contest (electorate size, voter turnout, majority size, rejected ballots, unreturned ballots), key demographic data for candidates (age, gender, ethnicity), and lineage data for political parties. MECo is the most well-curated open database on Malaysian elections to date, and will unlock new opportunities for research, data journalism, and civic engagement.
	\end{minipage}
\end{center}

\footnotetext[1]{Malaya University, W.P. Kuala Lumpur, Malaysia\\\Letter\space \href{mailto:thevesh.theva@gmail.com}{thevesh.theva@gmail.com}}

\vfill

\phantom{Word Count:}

\newpage
\pagenumbering{arabic}

\section*{Background \& Summary}
\begin{figure*}[htb]
	\centering
	\caption{Federal and state general election years}
	\vspace{0.14in}
	\makebox[\textwidth][c]{
		\includegraphics[width=1.1\linewidth]{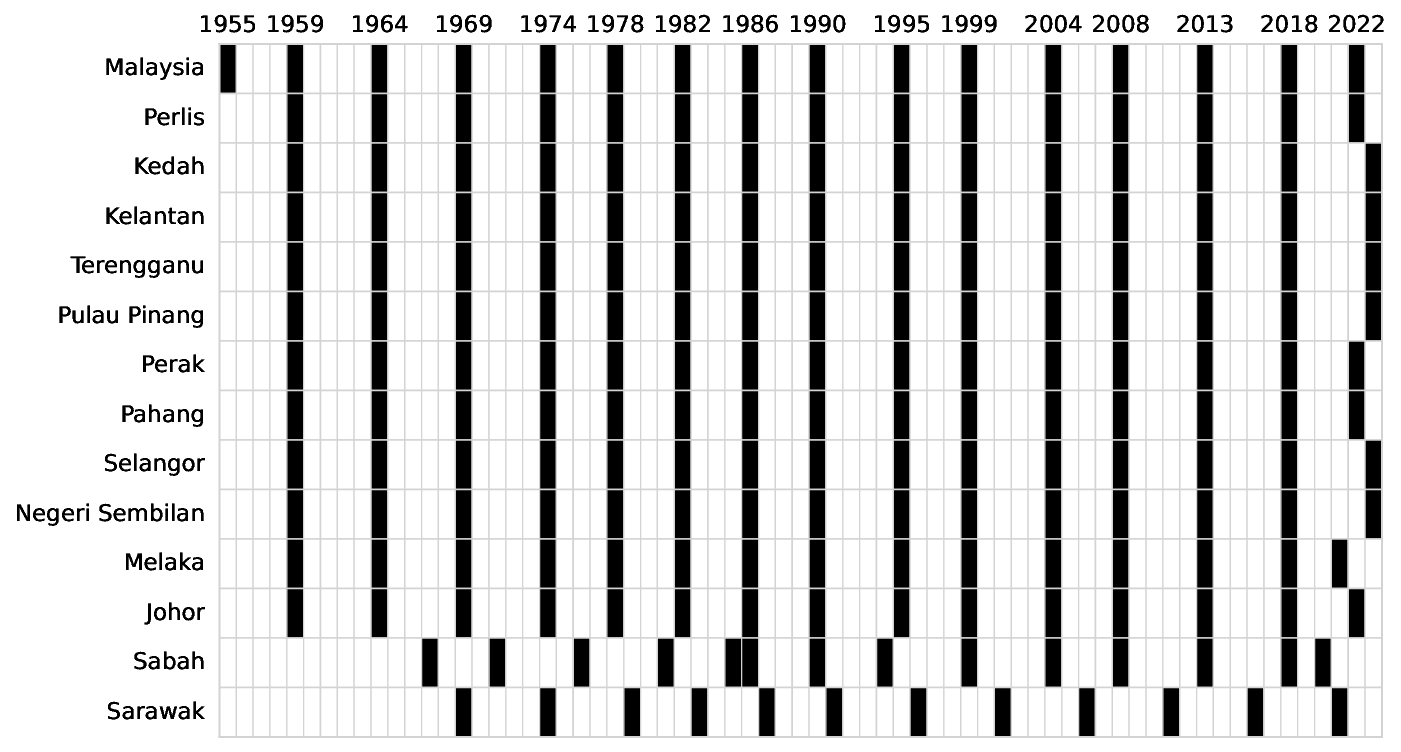}
	}
	\label{fig:elections}
\end{figure*}

Malaysia's electoral history is among the most dynamic in Southeast Asia, encompassing \TOTALCONTESTSFEDERAL\space federal election contests and \TOTALCONTESTSSTATE\space state election contests (Figure \ref{fig:elections}), as well as hundreds of off-cycle by-elections across a multiethnic, multi-party system. Furthermore, Malaysia offers significant scope for the study of democratisation, having experienced its first change of ruling party in 2018, and its first ever hung Parliament as recently as 2022. However, empirical studies of Malaysian elections are hindered by the lack of a comprehensive, standardised, and publicly available dataset that provides a single source of truth for scholars. The Election Commission (EC) does not publish open data which abides by best practices for data sharing, preferring instead to limit citizens to searching up isolated results via \href{https://citethis.link/myspr}{MySPR Semak}. Gazetted election results published as subsidiary legislation by the \href{https://citethis.link/warta}{Attorney General's Chambers (AGC)} contain slightly more detail, but are available in PDF format only. The lack of a Freedom of Information Act further complicates efforts to acquire and systematically compile electoral returns.

Amidst this paucity of data, global initiatives such as the Constituency-Level Elections Archive (CLEA)\cite{clea} provide immensely valuable cross-country coverage, including for Malaysian elections. However, they generally focus on federal contests, and within that scope, only on the number of votes won by each candidate (thus omitting information such as the electorate size, voter turnout, unreturned ballots, rejected ballots, etc). Similarly, international turnout or election integrity datasets\cite{icoma2023turnout,norris2014integrity} capture only high-level national indicators. Locally, a number of news and civil society organisations such as Tindak Malaysia\cite{tindak_data}, malaysiakini (\href{https://citethis.link/undinfo}{undi.info}), and Sinar Project compile election data to varying degrees of completeness and quality, but these efforts---while laudable for their public service, and valuable as a stopgap measure---typically lack proper data hygiene and standardisation, and are not subjected to systematic validation and review, thus limiting their usefulness for rigorous empirical research and long-term preservation.

In this paper, I address this gap by providing the Malaysian Election Corpus (MECo),\cite{meco_dataset} the first complete open database of Malaysian election results at the federal and state level. MECo is intended as a living resource which provides the go-to empirical foundation for research and journalism on Malaysian elections. The database covers \textit{all} general elections and by-elections since the pre-independence general election in 1955. In total, it records \UNIQUECANDIDATES\space unique candidates representing \TOTALPARTIES\space political parties in \TOTALCONTESTS\space unique electoral contests from 1955 to the present.

The dataset comprises two core components:
\begin{enumbul}
	\item \textbf{Ballots}: The final results for each state legislative assembly constituency (DUN) and federal Parliament constituency (Parliament), with the number of votes received by each candidate.
	\item \textbf{Stats}: The electorate size, ballots issued, unreturned ballots, and ballots rejected in each constituency. For each constituency, I also derive the margin of victory (majority), voter turnout rate, ballot rejection rate, unreturned ballot rate, and majority as a share of valid votes.
\end{enumbul}

Furthermore, my database offers three key advantages built on the use of unique identifiers (UIDs). First, I encode a UID for each candidate, allowing a single individual to be tracked across time even when they share a name with other candidates, contest under different parties, or change the name used in public life; this is especially important in Malaysia, where politicians are not required to use their official legal names on electoral ballots. Second, I encode a UID for each party, enabling consistent tracking of political parties even when they undergo rebranding or organisational transformation, such as the evolution of the National Justice Party (PKN) into the People's Justice Party (PKR) in 2003 following a merger with the Malaysian People's Party (PRM). Third, I encode a UID for each coalition, allowing coalition membership to be identified separately from party identity; this is an essential distinction in Malaysia, where election ballots frequently list the coalition rather than the actual party of the candidate. In general, the use of UIDs makes the database highly extensible, allowing other researchers to build new lookup tables or enrich existing ones without needing to alter the core datasets.

To the best of my knowledge, no comparable database exists. As a living resource, this paper lays the foundation for future data curation, as well as research into areas like malapportionment, gerrymandering, local-level voting patterns, and the spatial dynamics of political competition. I also hope that \acronym will serve as a catalyst for broader collaborations between academics, civil society, journalists, and election observers, supporting both scholarly inquiry and public accountability.

Finally, I note that while my work is the first of its kind for Malaysian elections, it follows a growing body of recent academic work focused on compiling and curating country-specific election data for reuse.\cite{perez2021spanish,baltz2022american,de2023american,calderon2025electoral,heddesheimer2025gerda,jensenius2017studying} By situating \acronym within this emerging tradition of high-quality electoral data curation, I contribute to the rapidly-improving global infrastructure of comparative political research. This work reflects a commitment to transparency, reproducibility, and the democratisation of access to electoral information.

\section*{Methods}
There are three main data sources I used to construct my database:
\begin{enumar}
	\item Gazetted election results published in PDF format by the \href{https://citethis.link/lom-pub}{AGC}; searching for ``results of contested election'' or ``keputusan pilihan raya yang dipertandingkan'' will yield the gazetted election results (Form 16, per Regulation 27, Electoral (Conduct of Elections) Regulations 1981). The downloadable PDF contains Malay and English versions of all information.
	\item Physical official election reports\cite{ge00report,ge01report,ge02report,ge03report,ge04report,ge05report,ge06report,ge07report,ge08report,ge09report,ge10report,ge11report,ge12report,ge13report,sabah1967report,sabah1971report,sabah1976report,sabah1981report,sabah1985report,sabah1986report,sabah1990report,sabah1994report,sabah1999report,sarawak1979report,sarawak1983report,sarawak1987report,sarawak1991report,sarawak1996report,sarawak2001report,sarawak2006report,kelantan1978report}.
	\item \href{https://citethis.link/myspr}{MySPR Semak}, an interactive website published by the EC; election results can only be queried one at a time, by selecting the appropriate election type (federal election, state election, or by-election), edition, and constituency. The site is only offered in Malay.
\end{enumar}
It should be noted that the first two sources are legally classified as open data and are not copyrightable under Malaysian copyright law. Section 3(1) of the Copyright Act 1987 (Act 332) expressly excludes from copyright “official texts of the Government or statutory bodies of a legislative or regulatory nature”. The gazetted election results and official election reports fall under this category as formal statutory publications. In the case of the interactive dashboard, the dashboard itself is copyrighted by the EC, but the underlying data is entirely derived from the gazetted results, and is therefore open data. This interpretation was confirmed through consultation with legal advisors and officials familiar with election-document provenance. Accordingly, I archived PDFs of the gazetted election results and post-election reports I used to construct \acronym (see Data Records).

I began with federal general elections, then state general elections, and finally by-elections. This is because state legislative assembly constituencies (DUNs) must lie completely within the boundaries of a federal parliamentary constituency (Parliament), so it was sensible to begin with the superset. By-elections coming last is an intuitive choice, since a by-election must follow a general election by definition; validation of by-election data is therefore dependent on having complete federal and state-level results.

\subsection*{Federal General Elections}

\begin{figure*}[htb]
	\centering
	\caption{Federal election coverage (number of seats)}
	\vspace{0.14in}
	\makebox[\textwidth][c]{
		\includegraphics[width=1.1\textwidth]{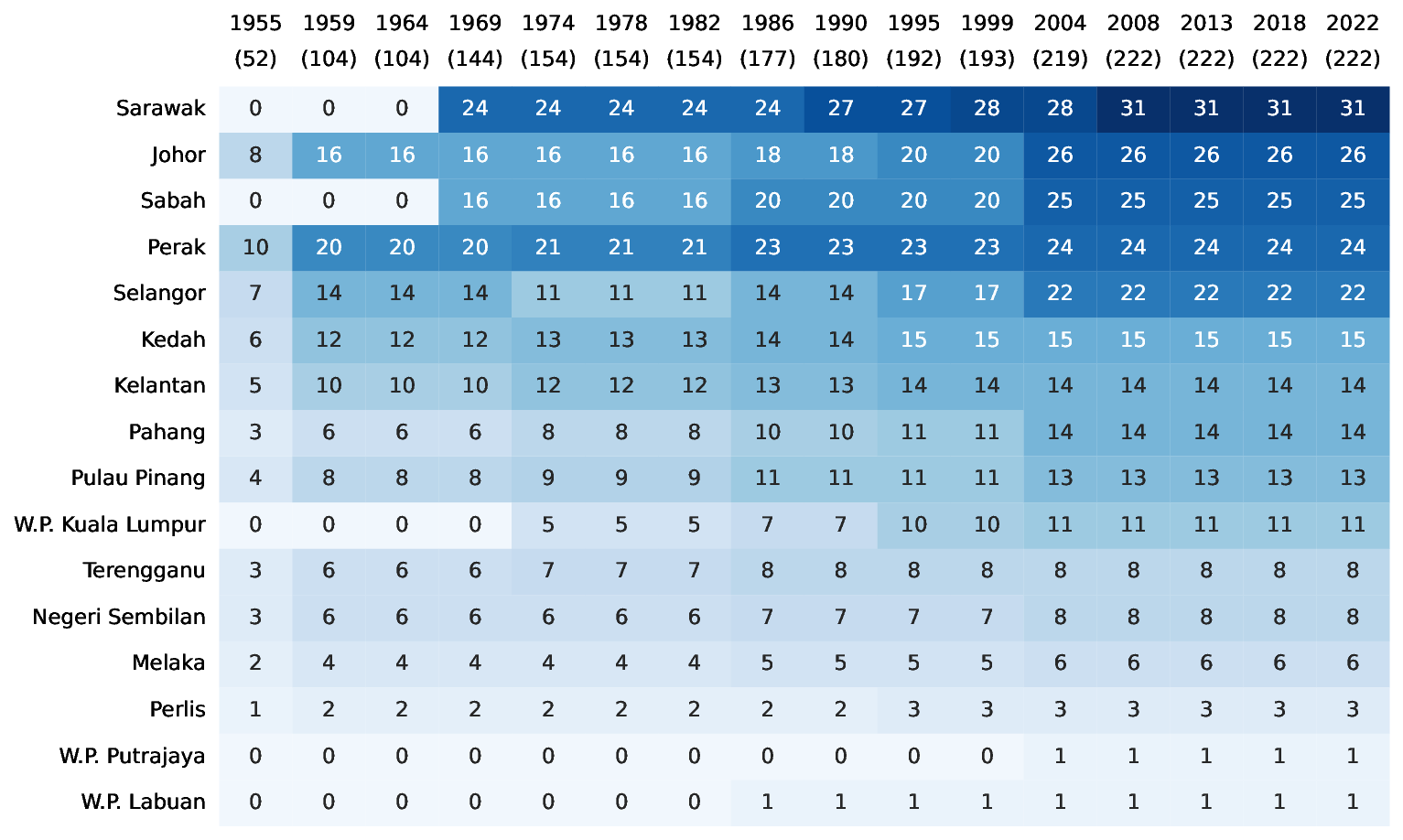}
	}
	\label{fig:federal_coverage}
\end{figure*}

For all federal elections in the dataset (Figure \ref{fig:federal_coverage}), I manually (i.e.\ by hand) digitised or copied the data from the aforementioned sources. I deliberately avoided the use of optical character recognition (OCR), PDF parsing, and web scraping tools after initial experimentation revealed an average error rate of approximately 10\%, primarily due to frequent changes in formatting and layout, even within the same document. I considered this to be unacceptably high for a resource intended to serve as a single source of truth. Moreover, the downstream process of error detection and correction proved to be more time-consuming and error-prone than simply transcribing the data by hand, especially given the relatively manageable size of this data (\TOTALBALLOTS\space rows of data across \TOTALCONTESTS\space electoral contests). The Technical Validation section further explains why the way in which Malaysia reports election results made it possible for me to do this with near-total confidence in the accuracy of the final data released for publication.

Records for seats in Peninsular Malaysia begin in 1955, while records for seats in Sabah and Sarawak begin in 1969. Although there was a federal general election in 1964, one year after Sabah and Sarawak (and Singapore) joined then-Malaya to form the Federation of Malaysia in 1963, seats in Sabah and Sarawak were not contested since the transition agreement allowed their respective state legislatures to appoint (and not elect) their representatives to the federal Parliament.\cite{means1991malaysia} There are no records for Singapore, which was not contested in the 1964 general election for the same reason as Sabah and Sarawak, and which exited the Federation prior to the next federal general election in 1969.

\subsection*{State General Elections}

\begin{figure*}[htb]
	\makebox[\textwidth][c]{
		\begin{minipage}{1.2\textwidth}
			\centering
			\caption{State election coverage (number of seats)}
			\vspace{0.14in}
			\includegraphics[width=1\linewidth]{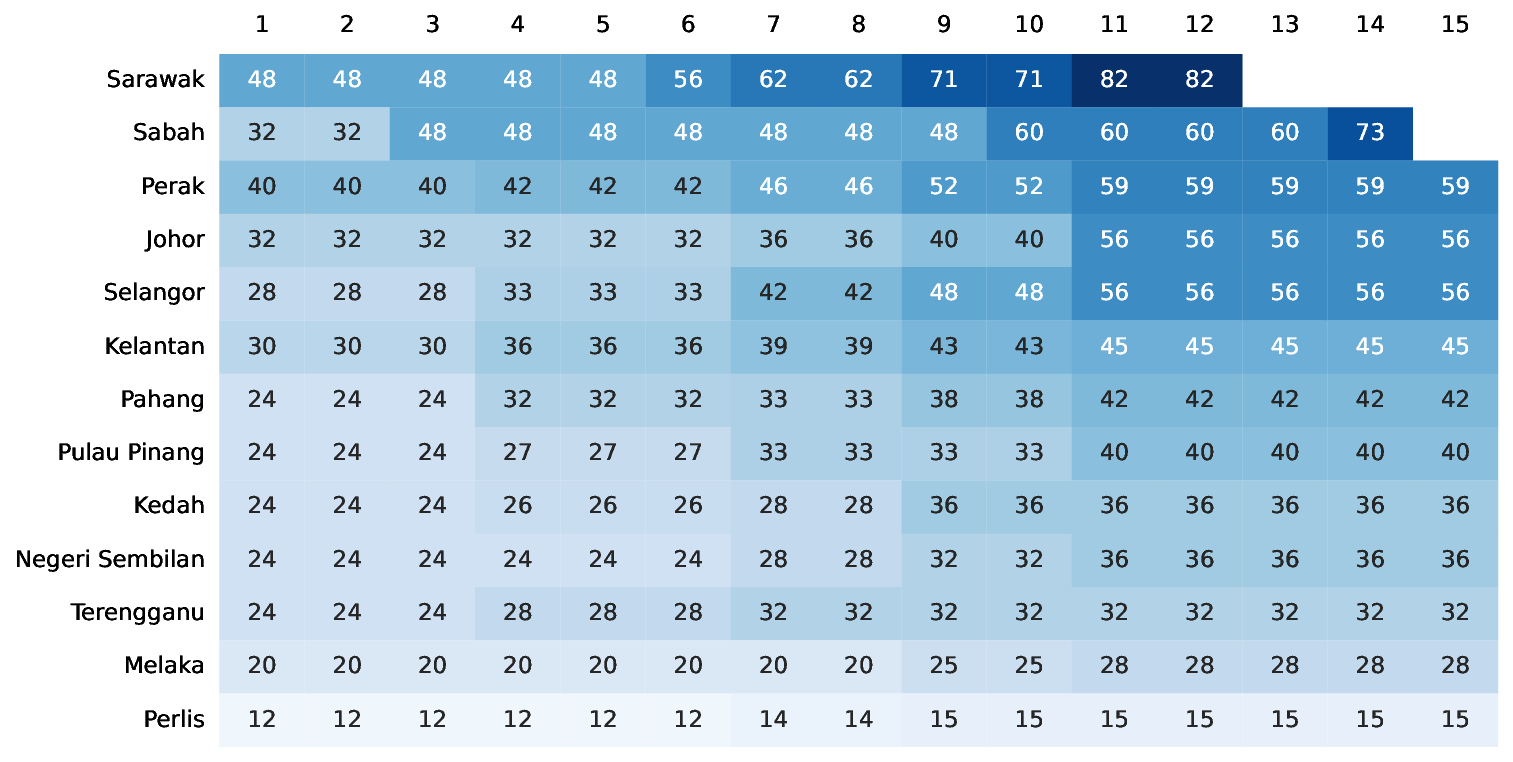}
			\label{fig:state_coverage}
		\end{minipage}
	}
\end{figure*}

After completing the federal general elections dataset, I constructed the state general elections dataset in exactly the same way.

For states in Peninsular Malaysia, records begin in 1959, when general elections for all 13 state legislative assemblies were held concurrently with the federal general election. For Sabah and Sarawak, records begin in 1967 and 1969 respectively, the years of the first state general elections held after the formation of the Federation of Malaysia. In all, \acronym contains records for 15 elections for all states in Peninsular Malaysia, 14 elections for Sabah, and 12 for Sarawak (Figure \ref{fig:state_coverage}). The reason for the discrepancy between the number of observations for Sabah and Sarawak is that there were two instances in Sabah's electoral history where state general elections were held in relatively quick succession. The first was in 1986, when Sabah went to the polls just one year after the previous state general election due to increasing civil and political instability\cite{yusoff2001sabah}. The second was in 2020, when Sabah held a state general election two years after the watershed election of 2018 due to a collapse of the state government.

\subsection*{By-Elections}

\begin{figure*}[htb]
	\centering
	\caption{By-election coverage (number of elections)}
	\vspace{0.14in}
	\includegraphics[width=0.8\linewidth]{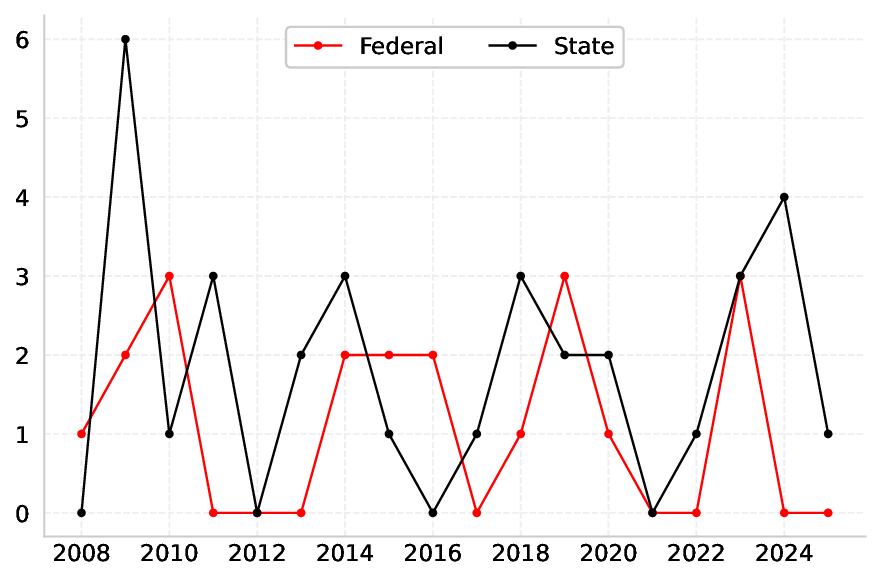}
	\label{fig:by_coverage}
\end{figure*}

Unlike for general elections, I could not locate any systematically compiled post-election reports or gazetted results for older by-elections. As a result, I relied solely on digital results made available via the EC's official website, which only covers by-elections since 2008. These digital records were copied, enriched, and standardised with the same schema used for general elections. Consequently, the by-elections dataset covers all Parliament and DUN by-elections held since the 12th federal general election in 2008, totaling 53 contests as of end-April 2025 (Figure \ref{fig:by_coverage}).

I plan to expand the dataset to include an exhaustive record of previous by-elections once I am able to acquire a reliable source, likely via a combination of EC reports and Hansards from Parliament and State Legislative Assemblies. However, given that the most recent three election cycles are already fully covered, the exclusion of older by-elections is likely negligible for almost all prospective users. I therefore made the decision not to delay the dissemination of \acronym any further.

\subsection*{Unique Identifier (UID) and Lookup Table Generation}

The database incorporates unique identifiers (UIDs) for parties, coalitions, and candidates to enable accurate longitudinal analysis. I also provide 3 lookup tables for parties, coalitions, and candidates, which allow users to augment the core datasets, either with variables I have already collated or with their own.

\textbf{Parties and Coalitions.}
Before discussing the generation of party and coalition UIDs, it is critical to distinguish how `party' is defined in this dataset relative to the EC's official records. When a candidate puts their name forth for nomination, they are mandated to declare themselves as either representing an entity registered with the EC, or as an independent candidate. However, the EC's \href{https://citethis.link/spr-parties}{list of registered parties} does not distinguish between a single political party and a coalition of parties. For example, as of 30th October 2025, the list contains---without distinction---both the United Malays National Organisation (UMNO) as well as the Barisan Nasional (BN). The former is a `party' in the conventional sense, while the latter is actually a coalition of several parties, including UMNO itself.

It is debatable as to whether the EC should amend their operating protocols to capture this distinction, or lower the barriers to a party being officially registered and greenlit for listing on a ballot. That judgment notwithstanding, the current set of practices has 3 particularly deleterious consequences for the quality of official election data:

\begin{enumar}
	\item \textbf{Misleading reporting of results for a single election}. Because the choice of party listed on the ballot is made at the candidate level, it is possible for candidates from the same party to list themselves differently. In practice, this happens because parties with a strong local identity choose to contest under the party flag within a particular state, but under the coalition flag in others. For example, the EC's official announcement of the 2022 general election results listed PN and PAS as having won 52 and 22 seats respectively. In actuality, PAS was a component party of PN, but chose to contest in Kelantan and Terengganu under the party flag due to its deep roots in those states. This greatly increases the likelihood of erroneous conclusions by users of official data, especially if they lack the required contextual knowledge.
	\item \textbf{Loss of insight into party dynamics}. Political parties which are part of stable coalitions often choose to contest under the coalitional identity. For example, UMNO---Malaysia's oldest political party---has never contested an election under its own banner, instead contesting under the Alliance flag until 1973, and the BN flag thereafter. This renders it impossible to analyse the political trajectory of UMNO relative to other component parties (especially MIC and MCA) using official data.
	\item \textbf{Loss of insight into unofficial coalitions}. For example, Pakatan Harapan was not officially registered in time for the 2018 general election, thus leading to candidates from its 3 component parties (PKR, DAP, and AMANAH) mostly choosing to contest under the PKR flag. While it is true that contemporaneous election observers were generally fully aware of this circumstance, future researchers may come to a different (incorrect) conclusion if they rely solely on official records.
\end{enumar}

For this reason, I made the methodological choice to capture 3 separate variables (see Data Records for complete schema). First, \texttt{party\_on\_ballot}---which faithfully captures the exact listing of the candidate on the official ballot. Second, \texttt{party}---which captures the true party allegiance of the candidate. Third, \texttt{coalition}---which captures the coalitional allegiance of the party. It was extremely difficult to consolidate this information going back to 1955, especially for state-level results which receive far less attention than federal results. However, I was able to eventually fill in all gaps using a combination of existing scholarly work,\cite{welsh2016elections, hai2017electoral, loh2002democracy, nohlen2001elections} media reports (for modern elections), archived news articles (for older elections), and consultation with colleagues specialising in election history. To the best of my knowledge, MECo is now the only publicly-available dataset with complete data allowing users to distinguish between the 3 dimensions of political allegiance captured here.

Once this was done, party and coalition UIDs were very easy to generate and validate manually, given that the total number of distinct parties (\TOTALPARTIES) and coalitions (\TOTALCOALITIONS) in the dataset is relatively small. The UIDs for coalitions are simple integers, while the UIDs for parties take the following form (e.g. \texttt{001-UMNO}):

\begin{centering}

	\texttt{\{integer\}-\{acronym\}}

\end{centering}

The reason I implemented this syntax is to encode sufficient information for `versioning' within the UID, given that instances of renaming and rebranding are fairly common in Malaysia's political landscape. For example, the Federated Sabah People's Front (UID: \texttt{065-BERSEKUTU}) was founded in 1994, renamed to the Sabah People's Front (\texttt{065-SPF}) in 2010, taken over and rebranded to the Sarawak Workers Party (\texttt{065-SWP}) in 2012, and then renamed again to the present-day Malaysian Nation Party (\texttt{065-PBM}) in 2021. An integer-only approach would have necessitated either `collapsing' these 4 iterations into one UID (resulting in loss of information), or using four different UIDs (resulting in loss of ability to chain versions). My solution enables users to detect that these 4 iterations are part of the same chain (via the 065 prefix), while still maintaining 4 separate rows in a lookup table so that information on each iteration can be provided. This syntax is also handy for distinguishing between parties using the same acronym; for instance, `UPKO' maps to 3 distinct parties which existed at different points in Sabah's election history.

Finally, for political parties, I used the UIDs as the basis for creating a separate lookup table tracking the lineage of political parties, i.e.\ noting down instances of merging, splitting, splintering, and renaming/rebranding linked to the UIDs of the predecessor and successor respectively. I did not deem this necessary for coalitions, which should be analysed by examining their party composition during elections, since coalition membership is always in flux and is independent of changes in the core identity of the coalition.

\textbf{Candidates.}
Candidate UIDs enable the accurate tracking of individual candidates across elections. Beyond the obvious use case of distinguishing candidates with the same name (e.g.\ the dataset contains 8 unique individuals with the exact same name---`Ahmad bin Abdullah'), robust candidate UIDs are particularly important in the Malaysian context, where candidates frequently acquire new honorifics, adopt different formatting conventions, or spell their name differently over time. Two illustrative examples demonstrate the need for candidate UIDs. 8-time Parliamentarian \textbf{Rafidah Aziz} appeared on ballots across 8 elections in forms including:

\begin{centering}

	Rafidah Aziz\\
	Rafidah Bt. Ab. Aziz\\
	Rafidah Bt. Abdul Aziz\\
	Datuk Seri Rafidah Aziz

\end{centering}

Another 8-time Parliamentarian \textbf{Samy Vellu A/L Sangalimuthu} appeared on ballots across 9 elections with variations such as:

\begin{centering}

	S. Samy Vellu\\
	Datuk Seri Samy Vellu\\
	S. Samy Vellu A/L Sangalimuthu\\
	Dato' S. Samy Vellu A/L Sangalimuthu

\end{centering}

Through careful examination, I assigned a single UID to all instances of each candidate, taking particular care to capture all permutations of an individual's name while distinguishing between different individuals who happened to share a name. Two cues were especially useful: \textit{space} and \textit{time}. Consecutive elections in the same seat often revealed consistent candidacy patterns (i.e.\ a `home base'), whereas large gaps across states or decades typically indicated distinct individuals despite identical names. These cues guided targeted searches of external information to ensure that each UID was assigned correctly. Prominent candidates were straightforward to verify, as anyone who won an election appears in public records. The most challenging cases were independent candidates who contested only once, especially in older elections prior to widespread digital documentation. The full process of assigning candidate UIDs took nearly a year; in MECo's present form, I am confident that most, if not all, detectable errors have been eliminated.

Once a complete and validated set of candidate UIDs was established, I created a candidate lookup mapping each UID to a cleaned name stripped of titles and honorifics. This lookup was then enriched with three demographic attributes: sex, ethnicity, and date of birth (DOB). Sex and ethnicity were assigned via manual inspection of names, which are nearly perfectly indicative of sex and strongly indicative of ethnic group in the Malaysian context; both variables are complete for 100\% of candidates. Where possible, these were verified against public records, particularly for recent elections. DOB were obtained through extensive manual searches across thousands of sources---including biographical directories, news archives, parliamentary profiles, and obituary notices---and recorded in ISO format (YYYY-MM-DD). Reliable birth information could not be found for 33.2\% of candidates overall, reflecting limited public documentation in earlier decades. Coverage improves markedly over time: 60\% of candidates in GE3 (1969) lack DOB information, falling to 16\% by GE10 (1999) and under 5\% by GE15 (2022), with remaining gaps concentrated among independent candidates with minimal public profiles.

\subsection*{No UID for Constituencies}

Constituency names are not reliable indicators of spatial continuity. The name of a constituency may remain the same despite substantial spatial change, as in the case of Lumut (Perak) in the 2018 delimitation exercise. Similarly, a constituency may be renamed despite the underlying territory remaining largely intact, as in the case of Silam (Sabah) being renamed to Lahad Datu in the 2019 delimitation exercise. More fundamentally, as several streams of work in the field have illustrated,\cite{hackett2025redrawing,crespin2005using,mcghee2017measuring} the notion of what makes a constituency the same or different across elections is inherently subjective, and should be calibrated to the specific research question at hand. Therefore, any attempt at generating a constituency UID using names alone would be a rough attempt at best, and incredibly misleading at worst.

Consequently, I consider constituency continuity best left to researchers. In fact, it arguably merits dedicated scholarly handling, since electoral lineage has never been rigorously documented or studied in the Malaysian context. That having been said, users who wish to augment \acronym with other constituency-based datasets---for example, the \href{https://citethis.link/opendosm-parlimen}{subnational statistics} published by the Department of Statistics Malaysia---should rely on the intrinsic uniqueness of the \texttt{date-state-constituency} combination as a composite key. This avoids imposing any particular lineage model, while ensuring that external data can be merged cleanly and reproducibly.

\section*{Data Records}

All datasets (Table~\ref{tab:data_records}) are published on Harvard Dataverse,\cite{meco_dataset} which serves as the canonical archive. For convenience, the exact same datasets are also mirrored on:

\begin{enumbul}
	\item \textbf{Zenodo}\cite{meco_codebase} (\href{https://doi.org/10.5281/zenodo.17694675}{doi.org/10.5281/zenodo.17694675})\\
	Provides code and raw source files, in addition to datasets.

	\item \textbf{GitHub} (\href{https://github.com/Thevesh/paper-malaysian-election-corpus}{github.com/Thevesh/paper-malaysian-election-corpus})\\
	Facilites active development and maintenance, issue tracking and community contributions. Substantial updates are released on Zenodo.
\end{enumbul}

The datasets fall into two groups; constituency-level statistics and candidate-level ballots, followed by lookup tables that extend the core schema.

\begin{table}[H]
	\centering
	\renewcommand{\arraystretch}{1.05}
	\caption{Description of primary datasets}
	\label{tab:data_records}
	\vspace{0.14in}
	\begin{tabular}{L{0.33\linewidth}L{0.67\linewidth}}
		\toprule
		\textbf{Filename}                  & \textbf{Description}                                                                                           \\
		\midrule
		\texttt{consol\_stats}             & Summary statistics for all federal and state elections                                                         \\
		\texttt{federal\_stats}            & Subset of \texttt{consol\_stats} for federal general elections                                                 \\
		\texttt{state\_***\_stats}         & Subset of \texttt{consol\_stats} for state general elections                                                   \\
		\texttt{byeelection\_stats}        & Subset of \texttt{consol\_stats} for by-elections                                                              \\[0.05in] \hline
		\texttt{consol\_ballots}           & Candidate-level results for all federal and state elections                                                    \\
		\texttt{federal\_ballots}          & Subset of \texttt{consol\_ballots} for federal general elections                                               \\
		\texttt{state\_***\_ballots}       & Subset of \texttt{consol\_ballots} for state general elections                                                 \\
		\texttt{byeelection\_ballots}      & Subset of \texttt{consol\_ballots} for by-elections                                                            \\[0.05in] \hline
		\texttt{lookup\_candidate}         & Standardised list of candidates                                                                                \\
		\texttt{lookup\_party}             & Standardised list of parties                                                                                   \\
		\texttt{lookup\_party\_succession} & Details of merging, splitting, splintering and rebranding                                                      \\
		\texttt{lookup\_coalition}         & Standardised list of coalitions                                                                                \\
		\texttt{lookup\_dates}             & Election dates, by state and election type                                                                     \\[0.05in]
		\hline
		\texttt{logs/corrections}          & Log of manual corrections applied to \texttt{ballots\_issued} as described in the Technical Validation section \\
		\bottomrule
	\end{tabular}
\end{table}

Table~\ref{tab:summary_structure} provides a detailed description of the \texttt{*\_stats} files. Each row corresponds to a single constituency in a single election, and contains the complete set of numerical aggregates for that contest. These files are therefore the best entry point for constituency-level analysis or for merging with external datasets organised at the constituency level. Users should note that the \texttt{*\_stats} files have a one-to-many relationship with the corresponding \texttt{*\_ballots} files; each row in a \texttt{*\_stats} file links to multiple candidate-level rows in the \texttt{*\_ballots} file, except in uncontested seats where only one candidate appears. The two file types can be joined using the composite key \texttt{(date, state, seat)}.

\begin{table}[H]
	\centering
	\renewcommand{\arraystretch}{1.3}
	\caption{Structure of all \texttt{*\_stats} files}
	\label{tab:summary_structure}
	\vspace{0.14in}
	\begin{tabular}{L{0.35\linewidth}L{0.65\linewidth}}
		\toprule
		\textbf{Variable}                     & \textbf{Description}                                 \\
		\midrule
		\texttt{date}                         & Date of the election (YYYY-MM-DD)                    \\
		\texttt{election}                     & Election name (e.g., GE-14, SE-10, BY-ELECTION)      \\
		\texttt{state}                        & State in which the seat is located                   \\
		\texttt{seat}                         & Full name of the seat (e.g., P.049 Tanjong)          \\
		\texttt{voters\_total}                & Total number of registered voters                    \\
		\texttt{ballots\_issued}              & Number of ballots issued                             \\
		\texttt{ballots\_not\_returned}       & Number of ballots not returned (often postal votes)  \\
		\texttt{votes\_rejected}              & Number of rejected (spoiled) ballots                 \\
		\texttt{votes\_valid}                 & Number of valid votes                                \\
		\texttt{majority}                     & Margin of victory (winner minus runner-up)           \\
		\texttt{n\_candidates}                & Number of candidates contesting                      \\
		\texttt{voter\_turnout}               & Ballots issued as a share of registered voters (\%)  \\
		\texttt{majority\_perc}               & Majority as a share of valid votes (\%)              \\
		\texttt{votes\_rejected\_perc}        & Rejected ballots as a share of ballots returned (\%) \\
		\texttt{ballots\_not\_returned\_perc} & Unreturned ballots as a share of ballots issued (\%) \\
		\bottomrule
	\end{tabular}
\end{table}

Table~\ref{tab:ballots_structure} provides a detailed description of the \texttt{*\_ballots} files. Each row corresponds to a single candidate contesting a specific constituency in a specific election. These files are therefore the best entry point for candidate-level analysis or for merging with external datasets organised at the candidate level. Party- or coalition-level analysis should also begin from these files.

\begin{table}[H]
	\centering
	\renewcommand{\arraystretch}{1.3}
	\caption{Structure of all \texttt{*\_ballots} files}
	\label{tab:ballots_structure}
	\vspace{0.14in}
	\begin{tabular}{L{0.29\linewidth}L{0.69\linewidth}}
		\toprule
		\textbf{Variable}          & \textbf{Description}                                    \\
		\midrule
		\texttt{date}              & Date of the election (YYYY-MM-DD)                       \\
		\texttt{election}          & Election name (e.g., GE-14, SE-10, BY-ELECTION)         \\
		\texttt{state}             & State in which the constituency is located              \\
		\texttt{seat}              & Code and full name of the seat (e.g., P.052 Bayan Baru) \\
		\texttt{ballot\_order}     & Order in which the candidate appeared on the ballot     \\
		\texttt{candidate\_uid}    & Unique identifier for the candidate                     \\
		\texttt{name\_on\_ballot}  & Candidate name as it appeared on the ballot paper       \\
		\texttt{party\_on\_ballot} & Party name as it appeared on the ballot paper           \\
		\texttt{party\_uid}        & Unique identifier for the party                         \\
		\texttt{party}             & True party allegiance of the candidate                  \\
		\texttt{coalition\_uid}    & Unique identifier for the coalition                     \\
		\texttt{coalition}         & True coalitional allegiance of the party                \\
		\texttt{votes}             & Number of valid votes received by the candidate         \\
		\texttt{votes\_perc}       & Share of valid votes received by the candidate (\%)     \\
		\texttt{rank}              & Rank of the candidate in that contest                   \\
		\texttt{result}            & Outcome (won, won uncontested, lost, lost deposit)      \\
		\bottomrule
	\end{tabular}
\end{table}

In addition to the primary ballots and statistics files, the database includes 5 lookup tables that support the core datasets. These lookup tables provide a structured way for users to perform deeper analysis or integrate external datasets without needing to modify the core data, thus guarding against downstream errors.

\begin{enumbul}
	\item \texttt{lookup\_candidate}: Standardised list of all candidates, including cleaned names (stripped of titles and honorifics) and demographic attributes (sex, ethnicity, and date of birth). This table should be left-joined onto any \texttt{*\_ballots} file using \texttt{candidate\_uid} as a key.

	\item \texttt{lookup\_party}: Standardised list of political parties, including party acronyms, full party names, and year of formation. This table should be left-joined onto any \texttt{*\_ballots} file using \texttt{party\_uid} as a key.

	\item \texttt{lookup\_party\_succession}: A lineage table documenting instances of parties merging, splitting, splintering, being absorbed, or rebranding. The \texttt{predecessor\_uid} and \texttt{successor\_uid} fields are fully consistent with the \texttt{party\_uid} field in the \texttt{lookup\_party} table, thus enabling joining in either direction as required.

	\item \texttt{lookup\_coalition}: Standardised list of electoral coalitions, including coalition UIDs, names, and short descriptions. This table should be left-joined onto any \texttt{*\_ballots} file using \texttt{coalition\_uid} as a key.

	\item \texttt{lookup\_dates}: A complete listing of all election dates for federal, state, and by-election contests. This table is provided to facilitate preparation of temporal joins with external datasets.
\end{enumbul}

\section*{Technical Validation}
There are five critical components of the database which require validation: Numerical data, candidate names, parties, coalitions, and constituencies.

For numerical data, the way in which Malaysia reports election results makes a unique form of validation possible. First, I define the following variables:
\begin{align*}
	I   & = \text{Ballots Issued}               \\
	U   & = \text{Unreturned Ballots}           \\
	R   & = \text{Ballots Rejected}             \\
	V_i & = \text{Valid Votes for Candidate } i
\end{align*}
For a given contest involving $N$ candidates, the following relationship must hold:
\begin{align}
	\label{eq:validation}
	I - U - R = \sum_{i=1}^{N} V_i
\end{align}
In plain language, the implied number of valid votes must be equal to the sum of votes received by all candidates. Because $I$, $U$, $R$, and $V_i$ are (and have been) reported separately in historical election reports and gazetted results, I can leverage this relationship as an almost-foolproof way to validate the accuracy of my manual data entry process. For transparency, I detected and corrected errors for 82 out of \TOTALCONTESTS\space contests, implying an error rate of 0.84\%, which is significantly lower than the 10\% error rate I encountered in my initial OCR experiments. Importantly, these errors are detectable and fixable, and come without the additional overhead of correcting errors in text data; my familiarity with local geography, politicians, and culture enabled me to transcribe text data rapidly and accurately.

This validation procedure also revealed \TOTALCORRECTIONS\space contests for which the data was digitised accurately based on my source material, but which nevertheless failed to satisfy Equation (\ref{eq:validation}). These errors presumably occurred due to mistakes in data entry which were not caught and corrected during the original publication process. In order to ensure a clean dataset, I applied a standard correction---for all \TOTALCORRECTIONS\space contests failing validation, I adjusted the number of ballots issued (as documented in \texttt{corrections.csv}) such that Equation (\ref{eq:validation}) holds. I chose \texttt{ballots\_issued} as the variable to adjust for two reasons. First, until the 1981 amendment of the Elections (Conduct of Elections) Regulations (implemented fully from the 1990 federal general election onwards), the number of unreturned ballots was not reported (and thus assumed to be 0). Second, because the number of rejected ballots is very small relative to the number of ballots issued, using rejected ballots as the adjusted variable would have resulted in a much larger correction in percentage terms---specifically, the correction would have required a 20\% change to the number of rejected ballots on average (with several exceeding 50\%), relative to a 0.6\% change to the number of ballots issued (with only one above 5\%, and none above 20\%).

\begin{figure*}[htb]
	\centering
	\caption{Error rate in general election years}
	\vspace{0.14in}
	\includegraphics[width=0.7\linewidth]{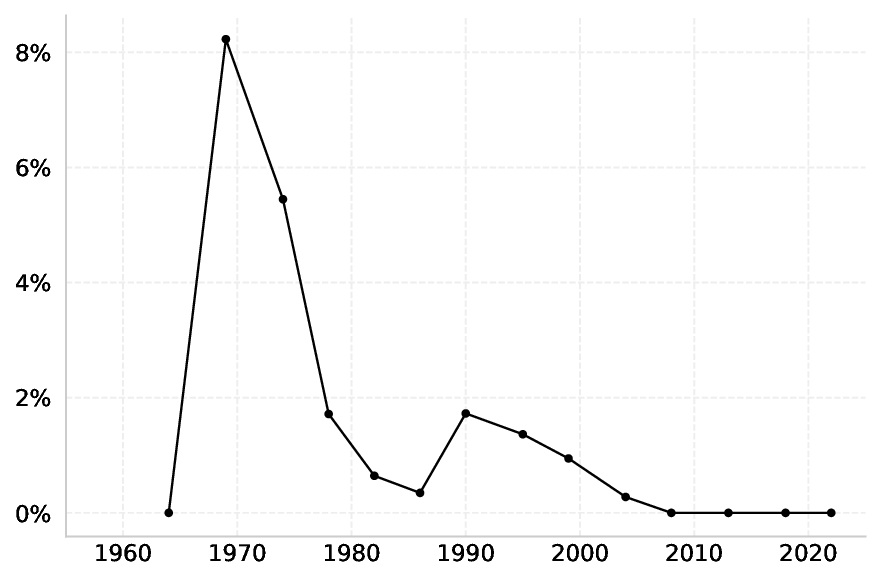}
	\label{fig:error_rate}
\end{figure*}

An interesting trend emerges when I plot the error rate arising from these \TOTALCORRECTIONS\space contests (Figure \ref{fig:error_rate}). The error rate was only 0.26\% in 1964 (just 1 error), but spiked to nearly 8\% in 1969, when elections were severely disrupted by the Emergency. By 1978, error rates stabilised below 2\%, and have been constant at 0\% since 2008. I posit that this reflects general improvements in data management technology over the decades; I do not have a clear explanation for the absence of errors in 1964.

\begin{figure*}[htb]
	\centering
	\caption{Histogram of derived variables}
	\vspace{0.14in}
	\includegraphics[width=0.8\linewidth]{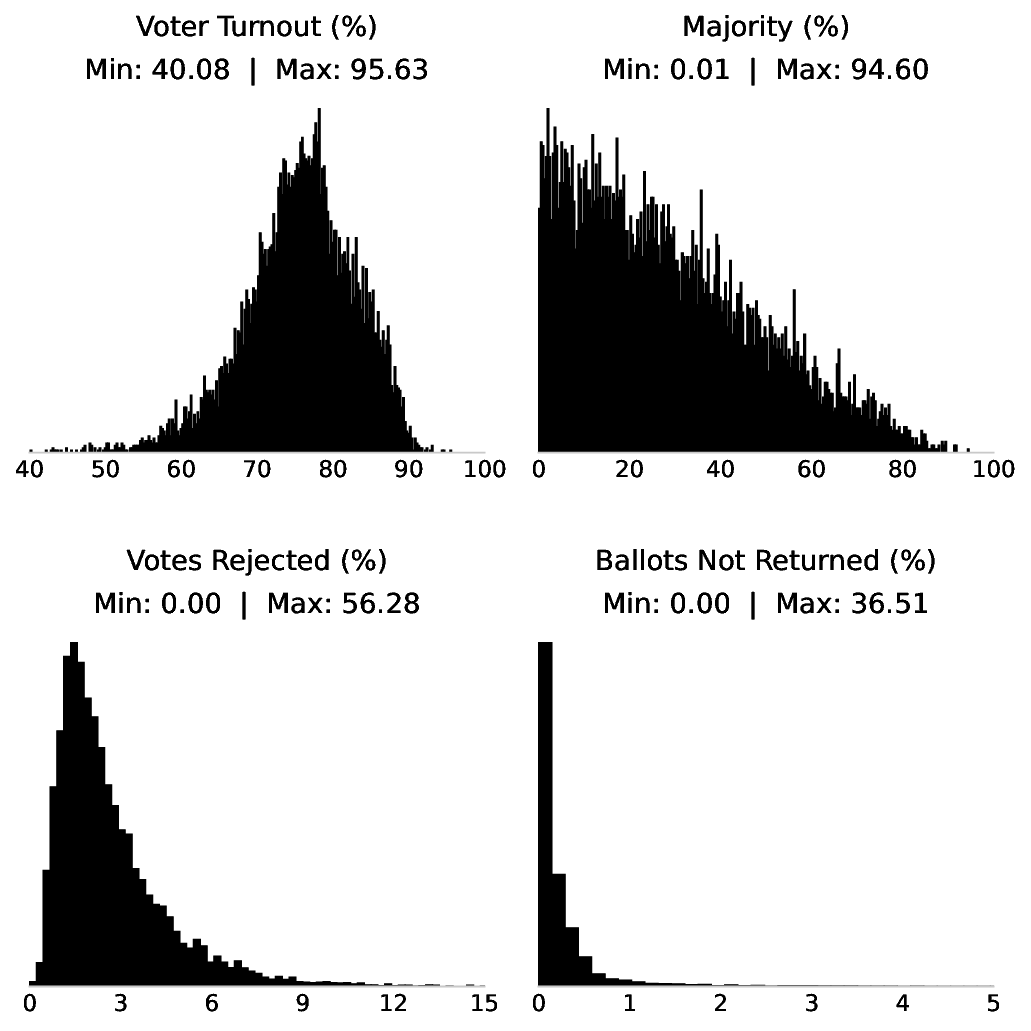}
	\label{fig:histograms}
\end{figure*}

Wrapping up my checks on numerical values, I note that the validation procedure can fail if two errors exactly offset each other. Furthermore, it cannot detect errors in the number of registered voters, which does not enter Equation (\ref{eq:validation}). To address these limitations, I plotted histograms of 4 derived variables (Figure \ref{fig:histograms}): voter turnout rate, ballot rejection rate, unreturned ballot rate, and majority as a share of valid votes. All four variables pass the check of being bounded between 0 and 100\%, and display smooth distributions as would be expected if data entry was accurate. All outlier values were double-checked; in particular, I verified that extreme instances of high rejected ballots or unreturned ballots were not due to mistakes in data entry. For example, the DUN of Pangkor in Perak had two instances where over 30\% of ballots were not returned; this was due to historical inefficiencies in the implementation of postal voting for navy personnel\cite{junid2019battle}.

For non-numerical data, I conducted 3 types of checks. For parties and coalitions, the full dataset only contains \TOTALPARTIES\space parties and \TOTALCOALITIONS\space coalitions, so I manually checked each against publicly available information. For constituency names, I generated a list of unique constituency names (1,437 in total), which I manually checked for errors in spelling or syntax. The most challenging component was candidate names and demographic details due to data volume (\TOTALBALLOTS\space rows). Although I applied rigorous formatting standards during data entry and conducted a full round of manual checking (line-by-line), I acknowledge that minor inconsistencies or inaccuracies may persist. This is especially because candidates often change the presentation of their names across elections (e.g., spelling out vs shortening a surname), acquire new titles and honorifics over time (e.g., Dato', Tan Sri, Haji, Ustaz, academic degrees), and format their names differently in different years (e.g., omitting or reordering titles). Such variations make consistent longitudinal tracking inherently complex. As with all living datasets of this scale, I anticipate incremental improvements over time as users engage with the data and identify potential refinements.

Finally, to ensure accuracy of the database as a whole, I derive seats and votes by party for all elections in the dataset, and ensure that these match against the record of parties in Parliament and State Legislative Assemblies.

\section*{Usage Notes}

The Malaysian Election Corpus (\acronymend) is designed to support a wide range of use cases, ranging from rigorous empirical research to rapid data journalism, and even casual civic technology projects. In the academic realm, \acronym---as the first database of its kind for Malaysia---serves as a foundational resource for research in electoral studies, political science, and public policy. Researchers can employ the data to answer important questions about the evolution of Malaysia's electoral system, employing both cross-sectional and longitudinal analysis.

The data can also be interactively explored via \href{https://electiondata.my}{ElectionData.MY}. In addition to making the data accessible to non-technical users, the site is intended to serve as the primary channel for continuous updates and enhancements. Future improvements to \acronym will be reflected on the site as new versions are archived on Harvard Dataverse\cite{meco_dataset} and Zenodo\cite{meco_codebase}, ensuring that users always have the freshest data.

I anticipate that many users will want to extend or adapt the dataset. The standardised schema and use of lookup tables enable seamless enrichment and integration with other datasets. Some valuable examples include:
\begin{enumbul}
	\item \texttt{lookup\_candidate} can be enriched with additional demographic or biographical information such as marital status, education, and occupation.
	\item \texttt{lookup\_party} and \texttt{lookup\_coalition} can be extended with variables such as ideological classification, membership size, or even beneficial ownership.
	\item Constituency-level data can be merged directly with \acronym using the intrinsic uniqueness of the \texttt{date-state-constituency} combination, thus enabling the use of \acronym in conjunction with official or alternative datasets.
\end{enumbul}
As a practical reference for users, the codebase contains samples (\texttt{dashboards.py}) of how to merge the core datasets with lookup tables to create panel data suitable for interactive dashboards and advanced analyses.

Finally, while rigorous validation has been applied (see Technical Validation), this remains a living database and is intended as such. Minor inaccuracies, particularly in the candidate name field, may persist as discussed above. I encourage users to report issues or submit improvements via the GitHub repository, which provides full transparency of changes between releases.

\section*{Data Availability}

All datasets described in this paper are published on Harvard Dataverse\cite{meco_dataset} under a CC0 license. For convenience, the exact same datasets are also mirrored in repositories on Zenodo\cite{meco_codebase} and \href{https://github.com/Thevesh/paper-malaysian-election-corpus}{GitHub}. Both contain code and raw source files in addition to the datasets; I use GitHub to version-control active development and maintenance, with substantial updates released on Zenodo.

\section*{Code Availability}

All data processing, validation, and compilation was conducted in Python. The full source code is publicly available under a CC0 license via Zenodo\cite{meco_codebase} and GitHub. Three key scripts which users should find particularly useful are:

\begin{enumbul}
	\item \texttt{compile.py}, which generates and validates all tabular datasets, with the exception of the \texttt{lookup*.csv} files, which were manually curated.
	\item \texttt{dataviz.py}, which generates all visualisations used in this paper.
	\item \texttt{dashboards.py}, which generates the source files for the interactive visualisations available at \href{https://electiondata.my}{ElectionData.MY}.
\end{enumbul}

% this section generates reference
\begingroup
\raggedright
\linespread{1.1}
\small
\let\oldthebibliography\thebibliography
\bibliography{mybib}
\endgroup

\section*{Acknowledgements}
I thank Rosmadi Fauzi and Zulkanain Abdul Rahman for their helpful feedback and suggestions on the dataset and manuscript. I also gratefully acknowledge Nimesha Thevananthan for her assistance with data collection, Danesh Prakash Chacko of Tindak Malaysia for providing historical insights and filling data gaps, and the MyElectionData volunteers for their assistance with completion and validation of candidate UID harmonisation and candidate demographic data. Finally, I thank the EC for granting access to physical election reports via \textit{Pusat Sumber SPR}.

\section*{Author Contributions (CRediT statement)}
TT: Conceptualization, Methodology, Software, Validation, Formal analysis, Investigation, Resources, Data curation, Writing - original draft, Writing - review and editing, Visualization, Supervision, Project administration.

\section*{Funding}

This work did not receive any specific funding.

\section*{Competing Interests}
The author declares no competing interests. The author is not affiliated with any political party or election-based organisation.

\end{document}